\documentclass[letterpaper,aastex]{emulateapj}
\usepackage{psfig}
\usepackage{apjfonts}

\shorttitle{Formation of UCXBs in  Dense Star Clusters}
\shortauthors{Ivanova et al.}

\begin{document}

\title{Formation of Ultracompact X-ray Binaries in Dense Star Clusters}

\author{N.\ Ivanova$^1$, F.~A.\ Rasio$^1$, J.~C.\ Lombardi$^2$, Jr.,
      K.~L.\ Dooley$^2$, and Z.~F.\ Proulx$^2$}

\altaffiltext{1}{Northwestern University, Dept of Physics and Astronomy,
Evanston, IL 60208.}
\altaffiltext{2}{Dept of Physics and Astronomy, Vassar College,
Poughkeepsie, NY 12604.}

\begin{abstract}{Bright, ultracompact X-ray binaries observed in dense
star clusters, such as Galactic globular clusters, must have formed
relatively recently, since their lifetimes as persistent bright sources
are short (e.g., $\sim 10^8\,$yr above $10^{36}\,{\rm erg}\,{\rm
s}^{-1}$
for a $1.4\,M_\odot$ neutron star accreting from a degenerate helium
companion with an initial mass of $\sim 0.2\,M_\odot$).
Therefore, we can use the present conditions in a cluster core to
study possible dynamical formation processes for these sources. Here we
show that direct physical collisions between neutron stars and red
giants can provide a sufficient formation rate to explain the observed numbers
of bright sources. These collisions
produce tight, eccentric neutron star -- white dwarf
binaries that decay to contact by gravitational radiation on
timescales $\sim 10^6-10^{10}\,$yr, usually shorter and often much
shorter than the cluster age.
}
\end{abstract}


\section{Introduction}

Ultracompact X-ray binaries (UCXBs) are persistent, bright X-ray sources
($L_x\sim 10^{36} - 10^{39}\,{\rm erg}\,{\rm s}^{-1}$) containing
a neutron star (NS) accreting from a low-mass, degenerate companion in
a very
tight orbit of period $P\la 1\,$hr. UCXBs may well be dominant among the
bright low-mass X-ray binaries (LMXBs) observed in old globular
clusters (GCs), both
Galactic \citep{DMA2000,2005A&A...431..647V} and
extragalactic \citep{BD2004}. It was recognized 30 years ago that
the total numbers of LMXBs observed in GCs
clearly indicate a dynamical origin, with formation rates exceeding 
those in
field
populations by several orders of magnitude (Clark 1975). Indeed, the
stellar
encounter rate in a cluster core is an
excellent predictor for the presence of a bright LMXB
\citep{2004ApJ...613..279J}.

The growing importance of UCXBs is clear from the role they have played
recently in a number of different contexts.
They may dominate the bright end of the X-ray
luminosity function in elliptical galaxies \citep{BD2004}.
They pose a number of challenges to, and may allow us to test our
fundamental physics of, stellar structure for low-mass degenerate or
quasi-degenerate objects \citep{DB2003}. 
They may also connect in a fundamental 
way to NS recycling, as suggested by the fact that three out
of six accretion-powered millisecond X-ray pulsars known in our Galaxy
are UCXBs \citep{C2004,IAUT2004}.
Finally, UCXBs may well be the progenitors of the many eclipsing binary
radio pulsars with very low-mass companions observed in GCs \citep{RPR2000}.

Several possible formation processes for UCXBs are possible.
Exchange interactions between NSs and primordial binaries provide
a natural way of forming possible progenitors of UCXBs \citep{RPR2000}.
This may well dominate the formation rate when integrated over the
entire history of a cluster. However, it is unlikely to be
significant for bright UCXBs observed {\it today\/}. This is
because the progenitors must be intermediate-mass binaries, with the NS
companion massive enough for the initial
mass transfer (MT) to become dynamically unstable, leading to common-envelope (CE)
evolution and significant orbital decay. Instead, all main-sequence
stars remaining today in a GC have masses low enough to lead to
{\it stable\/} MT (and orbits that expand during MT, 
leading to LMXBs with wide periods and non-degenerate donors).
Alternatively, some binaries with stable MT could evolve to
ultra-short periods by magnetic braking \citep{PS88,PRP2002}. 
However, producing UCXBs through this type of evolution requires very careful
tuning of initial conditions and is therefore very unlikely to
explain most sources \citep{vdSVP2004}.

\citet{V87} first proposed that a physical collision between a NS
     and a red giant (RG)
could lead to UCXB formation. In his scenario, the collision
was assumed to lead directly to a CE system in which the NS and
RG core would quickly inspiral. However,
RG--NS collisions that occur now in old GCs (where RGs
have low masses, close to $m_{\rm to}$) do {\it not\/} lead to
CE evolution. 
Instead, the RG envelope is promptly disrupted, leaving behind an eccentric NS--WD
binary, as shown by \citet{RS91} using 3-D hydrodynamic calculation.
Nevertheless, if the post-collision NS--WD binaries can
  decay through gravitational-wave emission all the way to contact,
they can still become UCXBs \citep{DBH92}.

\section{Outcome of Collisions}

Using the 3-D Smoothed Particle Hydrodynamics (SPH) code
{\tt StarCrash}\footnote{See
{\tt http://www.astro.northwestern.edu/StarCrash/.}} we have
computed about 40 representative collisions between various RG stars
and a $1.4\,M_\odot$ NS \citep{Letal2004}. In our models, both the NS
and the RG core are represented by point masses
coupled to the gas by (softened) gravity only \citep{RS91}.
Our initial RG models were calculated using the stellar
evolution code
described in detail in  \citet{PRP2002}, \citet{Iv03}, and \citet{Ka04}.
The models include stars on the subgiant branch
(with total mass $M=0.8M_\odot$, core mass $m_{\rm c} = 0.10\,M_\odot$
and radius $R_{\rm rg} = 1.6\,R_\odot$ and with $M=0.9M_\odot$,
$m_{\rm c} = 0.12\,M_\odot$ and $R_{\rm rg} = 2\,R_\odot$),
and several models near the base of the RG branch.
Our most evolved models have $M=0.9\,M_\odot$, $m_{\rm c} =
0.25\,M_\odot$
and $R_{\rm rg} = 6.8\,R_\odot$.
More evolved RGs contribute very
little to the total collision rate (see \S 3).
The distance of closest approach for the initial collision
varies from $r_p = 0.1\,R_{\rm rg}$ (nearly head-on) to
$r_p = 1.3\,R_{\rm rg}$ (grazing).

In agreement with previous SPH calculations \citep{RS91,DBH92},
we find that all
collisions produce bound systems in which the RG core
ends up in a high-eccentricity orbit around the NS.
However, in contrast to those older studies, our new SPH
calculations extend over much longer times (up to $\sim 500$
successive pericenter passages), allowing us to determine
accurately the final parameters of the orbit \citep{Letal2004}.
Typically $\sim50\%$ of the RG envelope is ejected to
infinity, while most of the rest becomes bound to the NS.
Only about $\sim 0.1\,M_\odot$ remains bound to the RG core,
which will eventually cool to a degenerate WD (cf.\ \S4).
     The material left bound to the NS will attempt to form an
accretion disk as it cools. The fate of this material is rather
uncertain. It could be accreted onto the NS and spin it up
(in $\sim 10^6\,$yr at the Eddington limit), or, more likely,
it could be ejected if the energy released by accretion couples
well to the gas. With an efficiency $\epsilon$,
the entire mass of gas could be ejected to infinity in as little as
$\tau_{\rm gas} \sim 500\,(\epsilon/0.1)^{-1}\,$yr.
This very short lifetime justifies our assumption that
the parameters of the post-collision orbits determined by our SPH
calculations are nearly final, i.e., that the orbital parameters
are no longer affected by coupling of the orbit to the residual gas.

When we apply the \citet{P64} equations to these post-collision systems,
we find that most of them inspiral on rather short timescales (Fig.\ 1).
Therefore, we assume for the rest of this paper that {\em all\/} RG--NS
collisions can produce UCXBs.

\begin{figure}
\centerline{\psfig{file=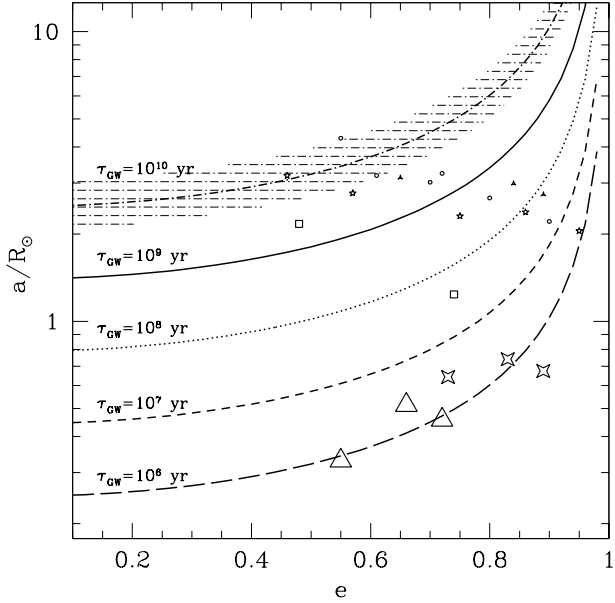,width=8.5cm}}
\caption{Dependence of the gravitational radiation merger time on
post-collision semimajor
axis $a$ and eccentricity $e$ for a binary consisting
of a $1.4\,M_\odot$ NS and a $0.25\,M_\odot$ WD.
The points with different symbols show the results of our SPH
calculations for six different giant models
(starred symbols are for a $0.9\,M_\odot $ star; others are for
a $0.8\,M_\odot$ star at different evolutionary stages,
e.g., triangles represent a subgiant, and circles our most evolved RG
model).
The symbol area is proportional to the collision
rate, according to eq.~(\ref{Rate_ucxb}), i.e., symbols for less 
evolved RGs
appear larger.
The hatched area shows how the merger time (here for the line of
constant
merger time $\tau_{\rm gw} = 10^{10}\,$yr) changes when we vary slightly
the binary
parameters: the upper boundary corresponds to a $1.5\,M_\odot$ NS with a
$0.45\,M_\odot$ WD, and the lower boundary corresponds to a
$1.3\,M_\odot$ NS with a $0.15\,M_\odot$ WD.
}
\end{figure}

\section{Collision Rate}

Consider a NS of mass $m_{\rm ns}$ in the core of a cluster containing
$N_\ast$ ordinary
stars (here we neglect binaries; see \S4). If all these ordinary
stars were turn-off stars of radius $R_{\rm to}$ and mass $m_{\rm to}$,
the collision rate for the NS would be
\begin{equation}
{\cal R}_{\rm to} \equiv {2\pi G (m_{\rm to}+m_{\rm ns}) N_\ast
R_{\rm to}}{\sigma}^{-1} {V_{\rm c}}^{-1},
\end{equation}
where $\sigma$ is the relative velocity dispersion,  $V_{\rm c}$ is the
core volume.
Here we assume that the collision cross section is
dominated by gravitational focusing 
.

To compute the collision rate with RGs, we take into account that the
number of RGs, $dN_{\rm rg}$, within any small range of radii
between $R_{\rm rg}$ and $R_{\rm rg}+ dR_{\rm rg}$ is
proportional to the time $dt$ spent there by the star as it ascends
the RG branch,
$dN_{\rm rg} = f_{\rm rg} N_\ast dt/\tau$.
Here $f_{\rm rg}$ is the fraction of stars with masses close enough to
the turn-off mass to have become RGs, and $\tau$ is the total lifetime
(from the
ZAMS to the end of the RG stage) of
a turn-off star, only slightly larger than the cluster age. For
a simple analytic estimate, we use the following approximate
relation between age and radius \citep[eq.~A9 of][]{KW96},
\begin{equation}
R_{\rm rg}(t)\simeq R_{\rm ZAMS} \left( 1 - \frac{t}{\tau}
\right)^{-0.28}, \label{R_rg}
\end{equation}
where we use $R_{\rm ZAMS}\simeq 0.7R_{\rm to}$. Next we replace $dt$
by $dR_{\rm rg}/(dR_{\rm rg}/dt) = (R_{\rm
to}/R_{\rm rg})^{4.6}
\,\tau\, dR_{\rm rg} / R_{\rm to}$.
The collision rate for a NS with RGs between
$R_{\rm rg}$ and $R_{\rm rg}+ dR_{\rm rg}$ is
\begin{equation}
d{\cal R} = {2.6 \pi G (m_{\rm to}+m_{\rm ns}) R_{\rm rg}
{\sigma}^{-1} {V_{\rm c}}^{-1}\, dN_{\rm rg}}\ . \label{dRate}
\end{equation}
Here a collision is defined to be any encounter with a distance of
closest
approach less than $1.3\,R_{\rm rg}$, consistent with our SPH results.
Integrating this over $R_{\rm rg}$ from the base of the RG branch,
defined by setting
$R_{\rm rg} \equiv b R_{\rm to}$, to the maximum radius of a RG,
$R_{\rm max} \gg b R_{\rm to}$, we find a total collision rate
\begin{equation}
{\cal R}_{\rm UCXB} \simeq 0.51 f_{\rm rg}\,
b^{-2.6}\, {\cal R}_{\rm to}.\label{Rate_ucxb}
\end{equation}
Alternatively, note that we could also directly integrate 
eq.~(\ref{dRate}) over
time, without changing the variable from $t$ to $R_{\rm rg}$. Because
the
collision rate is linearly proportional to radius when gravitational
focusing
dominates, we can then write
\begin{equation}
{\cal R}_{\rm UCXB} \simeq
{2.6 \pi G (m_{\rm to}+m_{\rm ns}) N_\ast f_{\rm rg} {\bar R}_{\rm
rg}}{\sigma}^{-1} {V_{\rm c}}^{-1}, \label{Rate}
\end{equation}
where ${\bar R}_{\rm rg}$ is the time-average radius of the RG. Using
eq.~(\ref{R_rg}) it is easy to show that eqs.~(\ref{Rate_ucxb}) 
and~(\ref{Rate}) agree.  Eq.~(\ref{Rate}) has the
advantage that any stellar evolution treatment can be used to determine
${\bar R}_{\rm rg}$, including fitting formulae more detailed than
eq.~(\ref{R_rg}) or numerical results from stellar evolution 
calculations.

The steep inverse dependence on $b$ in eq.~(\ref{Rate_ucxb}) indicates 
that the
collision rate
is completely dominated by the smallest RGs: although the cross
section increases
(linearly) with radius, the faster stellar evolution at larger radii
dominates, so that collisions are much more likely to happen when the
star is just leaving the main sequence, i.e., on or close to
the {\it subgiant\/} branch \citep{V87}. The corresponding core mass
is also small, typically $m_{\rm c}\simeq 0.1\,M_\odot$ for
$m_{\rm to}\simeq 0.8\,M_\odot$.

We now proceed to evaluate $f_{\rm rg}$. This depends on the mass
function
of stars in the cluster core, which we expect to be very different from
the IMF
because of mass segregation. Indeed, observations of cluster
cores reveal flat or even slightly rising mass functions
\citep[e.g.,][]{2004AJ....127.2771R}. Here we assume that the number of
stars within
$dm$ is proportional to $m^\alpha\,dm$, with $\alpha > -1$, between a
minimum $m_{\rm min}$ and a maximum $m_{\rm to} + \Delta
m$.
The spread $\Delta m$ of masses along the RG branch is obtained from the
mass-dependence of the main-sequence lifetime $t_{\rm ms}$. Adopting the
simple scaling $t_{\rm ms} = \tau (m_{\rm to}/m)^{3.6}$
\citep{2000MNRAS.315..543H}
we get
$\Delta t_{\rm ms} = 3.6\, \tau  (m_{\rm to} / m) ^{3.6}\, \Delta m
/m$. Setting
$\Delta t_{\rm ms} = t_{\rm rg}$, the total time spent on the RG branch,
and $m\simeq m_{\rm to}$ gives
$\Delta m \simeq 0.28 \, m_{\rm to} t_{\rm rg}/\tau.$

To be consistent with our previous definition of a RG having  a radius
$\ge b R_{\rm to}$ and using again eq.~(\ref{R_rg}) we derive
$t_{\rm rg}/\tau = 0.3 b^{-3.6}$. We can now calculate $f_{\rm rg}$ directly from IMF.
Assuming $\Delta m \ll m_{\rm to}$, and $m_{\rm min} \ll m_{\rm to}$
we get
\begin{equation}
f_{\rm rg} = (\alpha + 1) \frac{\Delta m}{m_{\rm to}}
               \simeq 0.28(\alpha+1) \frac{t_{\rm rg}}{\tau}
               = 0.08(\alpha+1) b^{-3.6}. \label{f_rg}
\end{equation}
Combining this with eq~(\ref{Rate_ucxb}) we obtain the result,
\begin{equation}
{\cal R}_{\rm UCXB} \simeq 0.04 \frac{\alpha+1}{b^{6.2}} \,{\cal
R}_{\rm to}.
\end{equation}

In steady state (justified given the short lifetimes $t_{\rm UCXB} \ll 
\tau$
of the bright UCXB phase) we can then estimate the number of UCXBs per 
$100$ NSs
at present in a cluster as
\begin{equation}
N_{100} \simeq 100 {\cal R}_{\rm UCXB} \, t_{\rm UCXB}.
\label{n_ucxb}
\end{equation}
The lifetime $t_{\rm UCXB}$ depends on the minimum luminosity for a
system to be classified as an UCXB.
For our estimates we adopt a minimum luminosity comparable
with
the observed minimum in our Galaxy,
$L_{\rm x} \simeq 10^{36}\,{\rm erg}\,{\rm s}^{-1}$.
The corresponding lifetime is $t_{\rm UCXB} \simeq 10^8\,$yr
\citep[e.g.,][]{RPR2000}.

The present mass of a cluster $M_{\rm t}$ is always less than its 
initial mass $M_{\rm t;0}= f_{\rm ML}^{-1} M_{\rm t}$, where $f_{\rm 
ML}$ is the total mass loss fraction.
About 40\% of the initial mass is lost just through stellar winds and 
SN explosions, so
that $f_{\rm ML} < 0.6$.
Without tidal mass loss (Joshi et al.\ 2001), and adopting a lower 
mass cut-off of $0.1\,M_\odot$ in the IMF of \citet{Kroupa_IMF_02}, 
we expect about 1 NS per $65\,M_\odot$ of mass at present
\citep[see also][]{2004astro.ph..5382I}.  About 5\% of these NSs will 
be retained,
depending on the escape velocity
and the NS natal kick velocity distribution \citep{2004astro.ph..5382I}.
The corresponding {\em minimum\/} number of UCXBs expected (without any 
tidal mass loss) is
then
\begin{equation}
N_{\rm min}\sim 8 \times 10^{-4} \, M_{\rm t} {\cal R}_{\rm UCXB} \, 
t_{\rm UCXB} ,
\end{equation}
where $M_{\rm t}$ is in $M_\odot$.

\begin{deluxetable}{ l l l l l l l l }
\tabletypesize{\scriptsize}
\tablecaption{UCXB formation in Galactic clusters.}
\tablehead{
\colhead{Cluster }  &
\colhead{  $t_{\rm rg}/\tau$}  &
\colhead{$\bar R_{\rm rg}$  }  &
\colhead{$\log\,\rho_0 $  }  &
\colhead{  $\sigma$}  &
\colhead{ $\log\, M_{\rm t}$ }  &
\colhead{ $N_{100} $ } &
\colhead{ $N_{\rm min} $ }
}
\startdata
NGC\,1851&     0.071&        6  &        5.7&        10.4&        6.0&  
    0.11 & 0.85\\
NGC\,6624&     0.087&        4.2&        5.6&        5.4&        5.2&   
    0.14 & 0.18\\
NGC\,6652&     0.076&        5.4&        4.8&        5.9&        5.4&   
    0.02 & 0.05\\
NGC\,6712&     0.070 &       5.9&       3.0&        4.3&        5.0&    
    0.0005 & 0.0004\\
NGC\,7078&     0.034&        7.1&        6.2&        12.0&        6.1&  
    0.16 & 1.62\\
Terzan\,5&     0.10&         4.3&        6.1&        10.6&        5.6&  
    0.27 & 0.87 \\
47\,Tuc  &     0.081&        4.9&        5.1&        11.5&        6.1&  
    0.23 & 0.23
\enddata
\label{table1}
\tablecomments{
The RG lifetime fraction $t_{\rm rg}/\tau$ and the average RG
radius $\bar R_{\rm rg}$ (in $R_\odot$) are calculated directly from
our stellar evolution code and
used in eq.~(\ref{f_rg}) (with $\alpha=0$) and eq.~(\ref{Rate}).
$\rho_0 $ is the cluster core
density (in $M_\odot\,$pc$^{-3}$), $\sigma$ is the (1-D) velocity
dispersion (in km\,s$^{-1}$) and $M_{\rm t} $ is the total cluster mass
(in $M_\odot$).
For Ter~5 and NGC 6652, $\log \rho_0$ is based on the
luminosity density from \citet{1993sdgc.proc..373D}
and an adopted mass-to-light ratio of 2. The value of $\log \rho_0$
for NGC 6652 appears rather uncertain \citep[see, 
e.g.,][]{1993sdgc.proc..357P, 1993sdgc.proc..373D}. Values of
$\log M_{\rm t}$ for NGC 6652 and Ter~5 are from
\citet{2002ApJ...568L..23G};
$\sigma$ for Ter~5 is from \citet{2002ApJ...568L..23G} and for NGC
6652 from \citet{1985IAUS..113..541W}. Otherwise $\log \rho_0$,  
$\sigma$ and $\log M_{\rm t}$
are from \citet{1993sdgc.proc..357P}.
}
\end{deluxetable}

In Table~1,  we show numerical results for several Galactic clusters:
all clusters where a UCXB has been identified,
and 47~Tuc (which does {\em not\/} contain any bright LMXB).
The probability of finding a bright UCXB in a cluster like 47~Tuc
is only about 23\%.
NGC 6652 has poorly measured parameters (see Note for Table~1),
and our numbers for this cluster are necessarily uncertain.
Two clusters, NGC~6624 and NGC~6712, are thought to have very
eccentric orbits and to be on the verge of complete
disruption in the Galactic tidal field \citep{1994A&A...290..412R,
1999ApJ...514..109G, 2001A&A...372..851A}.
This suggests that they may have had much higher mass and density in 
the past.
Indeed, observations show that  NGC~6712 has a strikingly unusual mass 
function
for stars below the turn-off
\citep{{2001A&A...372..851A}} and this can only be explained if the 
cluster
has lost more than 99\% of its initial mass \citep{2000ApJ...535..759T}
\footnote{After significant mass loss, and depending on its initial density 
profile,
the cluster could undergo strong gravothermal oscillations 
\citep{2000ApJ...535..759T},
so that a UCXB could also have formed when the core had a much higher 
density
during a recent, brief episode of core collapse.}.

\section{Discussion}

For Galactic clusters, our estimates indicate that it is quite possible
for all observed UCXBs to have been formed through RG--NS collisions 
(Table~1). For
extragalactic
clusters, we can crudely estimate the expected total number
of UCXBs in a galaxy by integrating over the cluster mass function and
assuming some average formation rate per NS in all clusters.
As an example, consider the case of M87 \citep{2004ApJ...613..279J}.
We adopt a power-law cluster IMF with $\alpha = -2$ \citep{Kravtsov}.
With an average formation rate ${\cal R}_{\rm UCXB} \sim  2 \times
10^{-12} - 4\times10^{-11}\,{\rm yr}^{-1}$ per NS (assuming
that M87 clusters have structural parameters distributed roughly
like those of Galactic clusters) and $t_{\rm UCXB} \sim 10^6\,$yr
(corresponding to $L_x > 10^{37}\,{\rm erg}\,{\rm s}^{-1}$,
near the detectability limit for M87), we find that $\sim 10-100$
UCXBs are expected in the 1688 identified clusters,
in rough agreement with the 58 detected LMXBs associated with these 
clusters.

In several galaxies, the probability of finding a bright LMXB in a
cluster appears
to correlate strongly with cluster metallicity
\citep{2002ApJ...574L...5K, 2003ApJ...589L..81K, 2004ApJ...613..279J}.
As there is no strong dependence of the RG--NS collision rate on
metallicity, in our scenario, we have to interpret
this trend as due to other factors, such as the metallicity dependence
of the IMF, of the number of NSs formed in the cluster, or of the NS 
retention
fraction.
In other words, the metallicity dependence must appear through
the number of NSs rather than through ${\cal R}_{\rm UCXB}$.
This is difficult to verify, since there are no well established 
theoretical or observational
predictions on how the IMF and NS natal kicks change with metallicity.
Alternatively, a strong metallicity dependence of $t_{\rm UCXB}$ is 
also possible,
with higher metallicity systems having longer lifetimes as bright
sources (cf.\ Maccarone et al.\ 2004).

Our conclusions are fairly robust, independent of assumptions and
in spite of some large theoretical uncertainties. We now examine
a few of the most important ones.
Binaries were neglected in our analysis. This implies that our
estimated collision rate is a lower limit, as interactions involving
binaries always {\em increase\/} this rate \citep{2004MNRAS.352....1F}.
However, the effects of binaries on collision rates in very dense
clusters {\em today\/} are likely to be small because core binary 
fractions
in these clusters are very small, typically a few percent at most.
This is known observationally \citep{2002scmc.conf..163C} and expected
theoretically (Fregeau et al.\ 2003; Ivanova et al.\ 2004b).
Another important assumption we made is that post-collision binaries
do not circularize. As seen in Fig.~1, high eccentricities are an
important factor in keeping merger times short. However, one can also
see directly from Fig.~1 that, even if {\em all\/} binaries were
able to circularize quickly (compared to the GR merger time), a large
fraction of post-collision systems would still merge in less than
the cluster age. Based on the results of Sec.~2 and the relation
between post-collision semimajor axis and collision parameters derived
from our SPH simulations, we estimate this fraction to be about 70\%.
Thus, even under the extreme assumption that all systems circularize,
the rate of UCXB formation would still be within a factor of 2 of
the total RG--NS collision rate.
One possible further complication could come from the residual gas
left bound to the RG core. All our collision calculations suggest that
the mass left bound to the RG core is $\sim 0.1\,M_\odot$. Although
there are many theoretical uncertainties, it is possible that this
is sufficient to reconstitute a RG envelope (Castellani et al.\ 1994).
In this case, the orbit would likely circularize, and {\em stable \/} MT 
from the reconstituted RG onto the NS would occur.
However, the Roche lobe in the post-collision binary is smaller than
the equilibrium radius of the RG, so that the MT
proceeds on a thermal timescale and the corresponding bright LMXB phase
lasts only $\sim 10^5\,$yr, making detection unlikely. In addition,
the total mass accreted by the NS will be only $\sim 10^{-3}\,M_\odot$,
which is not sufficient to produce a recycled millisecond pulsar.

\section*{Acknowledgments}
We thank R.\ Bi, S.\ Fleming, V.\ Kalogera, M.\ Rosenfeld and B.\
Willems for
helpful discussions. This work was supported by NSF Grants AST-0206276
and AST-0353997,
NASA Grants NAG5-12044 and NNG04G176G, and a Chandra Theory grant.


\ 

{\it Note added in proof.\/}
-- Bildsten \& Deloye (2004) pointed out that
the observed break around $5\times10^{38}\,{\rm erg}\,{\rm s}^{-1}$
in the X-ray luminosity functions of elliptical galaxies
could be explained naturally if all donors in UCXBs had initial
masses clustered near $\sim 0.1\,M_\odot$.
This is precisely what is predicted by our scenario.

\end{document}